%% file: main.tex
\documentclass[sigconf, nonacm, pdfa]{acmart}

\newcommand\vldbdoi{}
\newcommand\vldbpages{}
\newcommand\vldbvolume{}
\newcommand\vldbissue{}
\newcommand\vldbyear{}

\newcommand\vldbtitle{\shorttitle} 
\newcommand\vldbavailabilityurl{}
\newcommand\vldbpagestyle{empty} 

\usepackage{xcolor,listings}
\usepackage{textcomp}
\usepackage{color}

\definecolor{codegreen}{rgb}{0,0.6,0}
\definecolor{codegray}{rgb}{0.5,0.5,0.5}
\definecolor{codepurple}{HTML}{C42043}
\definecolor{backcolour}{HTML}{F2F2F2}
\definecolor{bookColor}{cmyk}{0,0,0,0.90}  
\color{bookColor}

\usepackage{algorithm}
\usepackage{algpseudocode}
\usepackage{amsmath}
\usepackage{bm}
\usepackage{multirow}
\usepackage{float}
\usepackage[section]{placeins}
\usepackage{subcaption}
\usepackage[normalem]{ulem}
\usepackage{balance}
\usepackage{appendix}
\usepackage[a-2b]{pdfx}

\usepackage{tabularx}

\begin{document}
\title{$ChatBI$: Towards Natural Language to Complex Business Intelligence SQL}

\author{
  Jinqing Lian{$^{\star}$},~~~Xinyi Liu{$^{\S}$},~~~Yingxia Shao{$^{\star*}$},~~~Yang Dong{$^{\S}$},~~~Ming Wang{$^{\S}$},Wei Zhang{$^{\S}$},\\~~~Tianqi Wan{$^{\S}$},~~~Ming Dong{$^{\S*}$},~~~Hailin Yan{$^{\S}$}
}
\affiliation{%
  \institution{
  {$^{\star}$} Beijing University of Posts and Telecommunications \\
  {$^{\S}$}Baidu Inc. \\        
  }
}
\email{{jinqinglian,shaoyx}@bupt.edu.cn, {liuxinyi02,dongyang06,wangming15,zhangwei143, wantianqi,dongming03,yanhailin}@baidu.com} 
\thanks{*Ming Dong and Yingxia Shao are the corresponding authors.}


\begin{abstract}

The Natural Language to SQL (NL2SQL) technology provides non-expert users who are unfamiliar with databases the opportunity to use SQL for data analysis.
Converting Natural Language to Business Intelligence (NL2BI) is a popular practical scenario for NL2SQL in actual production systems. 
Compared to NL2SQL, NL2BI introduces more challenges.

In this paper, we propose ChatBI, a comprehensive and efficient technology for solving the NL2BI task. 
First, we analyze the interaction mode, an important module where NL2SQL and NL2BI differ in use, and design a smaller and cheaper model to match this interaction mode. 
In BI scenarios, tables contain a huge number of columns, making it impossible for existing NL2SQL methods that rely on Large Language Models (LLMs) for schema linking to proceed due to token limitations. 
The higher proportion of ambiguous columns in BI scenarios also makes schema linking difficult. 
ChatBI combines existing view technology in the database community to first decompose the schema linking problem into a Single View Selection problem and then uses a smaller and cheaper machine learning model to select the single view with a significantly reduced number of columns. 
The columns of this single view are then passed as the required columns for schema linking into the LLM. 
Finally, ChatBI proposes a phased process flow different from existing process flows, which allows ChatBI to generate SQL containing complex semantics and comparison relations more accurately. 

We have deployed ChatBI on Baidu's data platform and integrated it into multiple product lines for large-scale production task evaluation. 
The obtained results highlight its superiority in practicality, versatility, and efficiency. 
At the same time, compared with the current mainstream NL2SQL technology under our real BI scenario data tables and queries, it also achieved the best results.

\end{abstract}

\maketitle

\pagestyle{\vldbpagestyle}
\begingroup\small\noindent\raggedright\textbf{PVLDB Reference Format:}\\
{XXX}. \vldbtitle. PVLDB, \vldbvolume(\vldbissue): \vldbpages, \vldbyear.\\
\href{https://doi.org/\vldbdoi}{doi:\vldbdoi}
\endgroup
\begingroup
\renewcommand\thefootnote{}\footnote{\noindent
This work is licensed under the Creative Commons BY-NC-ND 4.0 International License. Visit \url{https://creativecommons.org/licenses/by-nc-nd/4.0/} to view a copy of this license. For any use beyond those covered by this license, obtain permission by emailing \href{mailto:info@vldb.org}{info@vldb.org}. Copyright is held by the owner/author(s). Publication rights licensed to the VLDB Endowment. \\
\raggedright Proceedings of the VLDB Endowment, Vol. \vldbvolume, No. \vldbissue\ %
ISSN 2150-8097. \\
\href{https://doi.org/\vldbdoi}{doi:\vldbdoi} \\
}\addtocounter{footnote}{-1}\endgroup

\ifdefempty{\vldbavailabilityurl}{}{
\vspace{.3cm}
\begingroup\small\noindent\raggedright\textbf{PVLDB Artifact Availability:}\\
The source code, data, and/or other artifacts have been made available at \url{\vldbavailabilityurl}.
\endgroup
}

\input{1.intro}

\input{2.background}

\input{3.tech}

\input{5.expr}

\input{6.related_conclusion}

\begin{acks}
The authors are grateful to the Baidu MEG's MEDD team, and in particular, to Jiawei Liu, Chaoxian Gui, Yuchao Jiang, and many fellow team members. 
\end{acks}


\balance
\bibliographystyle{ACM-Reference-Format}
\bibliography{sample}

\end{document}

%% file: 1.intro.tex
\section{Introduction}

The rapid development of LLMs has attracted the attention of researchers in both academia and industry, with researchers from the Natural Language Processing (NLP) and database communities hoping to use LLMs to solve the NL2SQL task. 
The NL2SQL task involves converting user Natural Language into SQL that can be executed by a database, with the SQL semantics being consistent with the user's intent. 
Today, thousands of organizations, such as Google, Microsoft, Amazon, Meta, Oracle, Snowflake, Databricks, Baidu, and Alibaba, use Business Intelligence (BI) for decision support. 
Consequently, researchers representing the industry have quickly focused on the most attention-grabbing scenario of the NL2SQL task in actual production systems: the NL2BI task, which involves converting Natural Language into BI through technology. 
Through NL2BI technology, many non-expert users who do not understand databases, such as product managers or operations personnel, can perform data analysis, thereby aiding decision-making. 

\begin{figure*}[t]
  \includegraphics[width=0.985\textwidth]{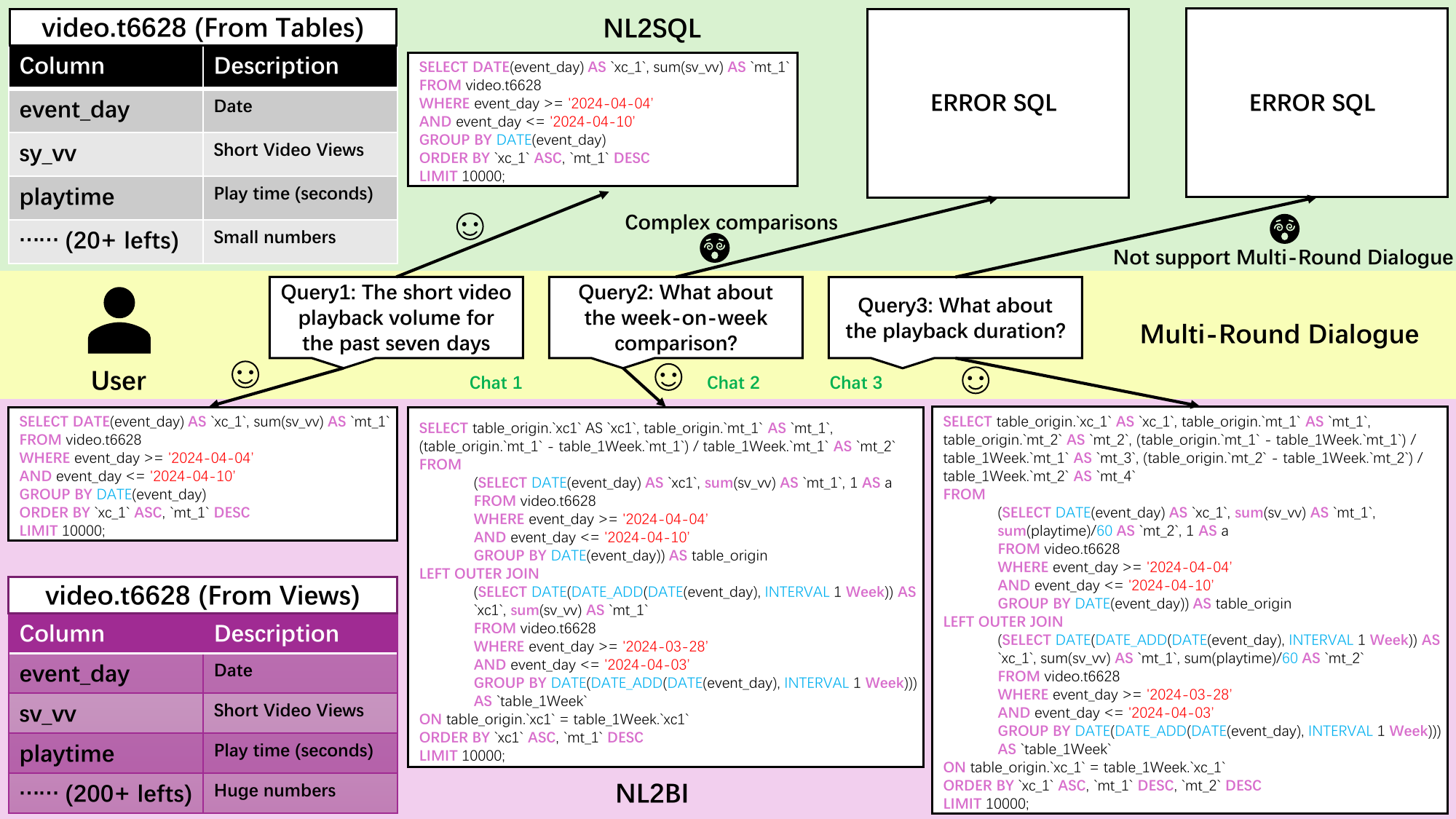}
  \caption{The difference between NL2SQL and NL2BI problems. The question consists of multiple rounds of dialogue, Query1: ``\textit{The short video playback volume for the past seven days}'', Query2: ``\textit{What about the week-on-week comparison?}'', Query3: ``\textit{What about the playback duration?}''. The NL2BI problem requires the ability to handle complex semantics, comparisons and calculation relationships, as well as multi-round dialogue. And the data sources for the two problems are also different. }
  \Description{The BI example.}
  \label{figure:BIExample}
\end{figure*}

Figure~\ref{figure:BIExample} reflects the difference between the NL2BI and NL2SQL tasks. 
In actual production systems, it is humans who call the NL2BI tool, and the most significant characteristic of humans using the tool is interactivity, i.e.,  Multi-Round Dialogue (MRD) scenarios. 
For example, the three-round dialogue scenario shown in Figure~\ref{figure:BIExample}. 
The Query1 indicates ``\textit{the user's query for the column of short video playback volume over the past seven days}''. 
Both NL2SQL and NL2BI technologies can recognize the user's intent and generate the corresponding executable correct SQL. 
In the Query2, the user asks about the common query in BI scenarios, ``\textit{the Week-on-Week comparison}'', which corresponds to complex semantics and comparison relationships. 
The presence of complex semantics and comparison relationships makes the problem of generating SQL difficult, and existing NL2SQL technology cannot generate the SQL required for BI well. 

Existing LLMs-based NL2SQL technology only matches Single-Round Dialogue (SRD) queries, and when there is no prompt to show that the next question is a MRD, the LLM cannot recognize the intent of the MRD well. 
Therefore, for the Query3 changing the intent of the column, the existing NL2SQL method cannot recognize that the user's intent is to query ``\textit{the playback duration of short videos over the past seven days and the Week-on-Week comparison}''. 
Therefore, the existing NL2SQL method still cannot generate the SQL corresponding to the intent of the Query3. 
In addition to the different interaction modes brought by actual scenarios, data tables under BI scenarios often contain hundreds of columns, most of which exist in views, while tables in NL2SQL usually contain dozens of columns, most of which come from real physical tables~\cite{floratou2024nl2sql}. 

Researchers have not put much effort into NL2BI, but they have invested considerable effort in NL2SQL. 
Existing methods can be divided into three categories. 
The first category is pre-trained and Supervised Fine-Tuning (SFT) methods~\cite{sutskever2014sequence, graves2012long, choi2021ryansql, hwang2019comprehensive, wang2019rat, cai2021sadga, cao2021lgesql, gan2021natural, yin2020tabert, herzig2020tapas, yu2020grappa}, which fine-tune an ``encoder-decoder'' model and use a sequence-to-sequence pattern to convert Natural Language into SQL. 
This was the mainstream approach before the advent of LLMs. 
The second category is prompt engineering based LLMs. 
Inspired by chain-of-thought~\cite{wei2022chain} and least-to-most~\cite{zhou2022least}, more and more research focuses on designing effective prompt engineering to make LLMs perform better in various subtasks. 
For example, in the NL2SQL scenario, DIN-SQL~\cite{pourreza2024din}, C3~\cite{dong2023c3}, and SQL-PaLM~\cite{sun2023sql} provide prompt engineering to GPT-4~\cite{openAIGPT}, GPT-3.5~\cite{openAIGPT}, and PaLM~\cite{narang2022pathways}, respectively, to greatly enhance the accuracy of generating SQL from Natural Language. 
The third category is LLMs specifically trained for NL2SQL, which use domain knowledge in NL2SQL to train specialized LLMs. 
However, when directly applying these methods to real analysis tasks in the BI scenario, we encountered some problems from the following perspectives:

\noindent
\textbf{C1: Limited interaction modes.}
The MRD interaction scenario shown in Figure~\ref{figure:BIExample} considers the interactivity of users using the NL2BI method, while the vast majority of existing NL2SQL methods, due to the evaluation of the open-source leaderboards Spider~\cite{yu2018spider} and BIRD~\cite{li2024can}, conduct queries in a SRD mode, neglecting MRD scenarios and limiting user interaction modes. 
For example, for the prompt engineering based LLMs such as DIN-SQL~\cite{pourreza2024din}, MAC-SQL~\cite{wang2023mac}, and DAIL-SQL~\cite{gao2023text}, the prompt engineering is organized in a SRD mode. 
Simply adding MRD prompts to the prompt engineering does not match MRD scenarios. 
The most challenging aspect is that the LLM needs to first determine whether a given sentence is part of a MRD or a SRD. 
For example, Query1 can be considered the SRD mode because it can generate SQL that corresponds to the intended meaning. 
However, Query2 cannot be considered the SRD mode, and it requires combination with Query1 to generate SQL that meets the intended meaning. 
Therefore, there is a need for an accurate MRD matching method that can satisfy the user interactivity.

\begin{table}[t]
\caption{Examples of ambiguous columns in BIRD dataset.}
\label{table:BIRDExample}
\begin{tabular}{lll}
\hline
db\_id             & column\_name  & description                \\ \hline
ice\_hockey\_draft & sum\_7yr\_GP  & sum 7-year game plays      \\
ice\_hockey\_draft & sum\_7yr\_TOI & sum 7-year time on ice     \\
hockey             & hOTL          & Home overtime losses       \\
hockey             & PPG           & Power play goals           \\
language\_corpus   & w1st          & word id of the first word  \\
language\_corpus   & w2nd          & word id of the second word \\ \hline
\end{tabular}
\end{table}

\noindent
\textbf{C2: Large number of columns and ambiguous columns.}
In the NL2BI scenario, SQL mainly consists of Online Analytical Processing (OLAP) queries. 
From a timeliness perspective, this leads to tables in BI scenarios often undergoing Extract-Transform-Load (ETL) processing first, incorporating a large number of columns into the same table to avoid join queries. 
From the perspective of actual storage, this new table containing a large number of columns often exists in the form of views. 
For example, one of Microsoft's internal financial data warehouses consists of 632 tables containing more than 4000 columns and 200 views with more than 7400 columns. 
Similarly, another commercial dataset tracking product usage consists of 2281 tables containing a total of 65000 columns and 1600 views with almost 50000 columns~\cite{floratou2024nl2sql}. 
The large number of columns poses a direct challenge to methods that rely on LLMs for schema linking, such as DIN-SQL~\cite{pourreza2024din}, MAC-SQL~\cite{wang2023mac}, and DAIL-SQL~\cite{gao2023text}. 
When columns are input as tokens, it causes the number of tokens to exceed the limit, and simply truncating columns leads to a sharp drop in the accuracy of generated SQL~\cite{floratou2024nl2sql}. 
At the same time, the proportion of ambiguous columns in BI scenarios has also expanded. 
Table~\ref{table:BIRDExample} shows the columns ambiguity in the BIRD dataset, which accounts for a small proportion. 
However, in BI scenarios, column ambiguity accounts for a larger proportion. 
For example, the ``uv'' can be represented by the ``User view'', ``Unique visitor numbers'', and ``Distinct number of visits''. 
The presence of column ambiguity will increase the likelihood of hallucinations in LLMs.

\noindent
\textbf{C3: Insufficiency of the existing process flows.}
The existing process flows use LLMs to directly generate SQL. 
For example, MAC-SQL~\cite{wang2023mac} shows accuracies of 65.73\%, 52.69\%, and 40.28\% for \textbf{Simple}, \textbf{Moderate}, and \textbf{Challenging} tasks, respectively, on the BIRD dataset. 
DAIL-SQL~\cite{gao2023text} also has similar results. 
However, SQL tasks in BI scenarios often contain complex semantics, comparisons~\footnote{Week-on-week comparisons in Figure~\ref{figure:BIExample}.}, and calculation relationships~\footnote{(table\_origin.`mt\_1' - table\_1Week.`mt\_1') / table\_1Week.`mt\_1' AS `mt\_2' in Figure~\ref{figure:BIExample}.}, making the tasks usually \textbf{Challenging+}. 
Advanced LLMs demonstrate superior performance on NL2SQL tasks~\cite{floratou2024nl2sql}, albeit with higher costs (GPT-4-32K is about 60X more expensive than GPT-3.5-turbo-instruct~\cite{oepnAIPricing} at the time of this writing). 
Most current process flows employ advanced LLMs such as GPT-4~\cite{openAIGPT}.
However, GPT-4 has not yet opened an entry for SFT, limiting existing approaches to handling complex semantics, computations, and comparison relationships through prompt. 
Longer prompts tend to yield poorer results~\cite{floratou2024nl2sql}, significantly constraining the understanding of the aforementioned challenges within the existing process flows.
This means that existing LLMs-based NL2SQL methods cannot effectively handle complex semantics, calculations, and comparison relationships in BI SQL. 
This reflects the insufficiency of the existing process flows, where all existing methods use LLMs to understand challenging complex semantics, comparisons, and calculation relationships. 
In BI scenarios, calculation relationships are easy to list (such as Week-on-Week and Year-on-Year comparisons) but difficult for LLMs to directly perceive. 
Therefore, there is an opportunity to find a new process flow that can better handle complex semantics, comparisons, and calculation relationships.

\noindent
\textbf{Our approach.} 
To address the challenges described above, we introduce ChatBI for solving the NL2BI task within actual production systems. 
For \textbf{C1}, unlike previous methods that focus solely on SRD scenarios, we treat the matching of MRD scenarios as a crucial module and identifies corresponding features, \textit{columns}, and \textit{dimensions}. 
We employ two smaller and cheaper Bert-like models to first conduct \textit{text classification} and then \textit{text prediction} for MRD matching. 
Compared to the method of inputting all dialogues as a sequence for prediction, ChatBI introduces a \textit{recent matching mode}, which shows improved performance. 
For \textbf{C2}, we initially integrate mature view technology from the database community, converting the schema linking problem into a \textbf{Single View Selection problem}. 
A smaller and cheaper single view selection model is used to select a single view, and all columns within this view are treated as the schema. 
The ability of views to retain a single column for ambiguous columns effectively addresses the problem of column ambiguity. 
Lastly, for \textbf{C3}, since the existing process flows predominantly utilize GPT-4, which cannot perform SFT and must rely solely on prompts, leading to a failure to understand complex semantics, comparisons, and computational relationships, we have designed a phased process flow and introduced virtual columns. 
Using this innovative process flow and virtual columns, we decouple LLMs from the understanding of complex semantics, comparisons, and computational relationships, and SQL language itself, thereby increasing their accuracy by reducing the task difficulty faced by the LLMs. 
In the new process flow, we position the SQL generation step after the LLM using existing rule-based methods, thus avoiding most errors caused by the hallucinations of LLMs.

Specifically, we make the following contributions:

\begin{itemize}
\item {} Considering the MRD scenario that objectively exists in the real-world use of NL2BL technology, we propose and solve the MRD problem in the NL2BI scenario. 
\item {} Unlike previous methods that directly use LLMs for schema linking, we transform schema linking into a single view selection problem using view technology from the database community. 
We then use the columns corresponding to the single view as the selected columns input to the LLM. 
\item {} Unlike previous methods that directly use LLMs to generate SQL, we propose a phased process flow. This process outputs structured intermediate results, decoupling some content that the LLM needs to perceive from the original process. Complex semantics, calculations, and comparison relationships are handled through structured intermediate results using mapping and other methods. The structured nature of SQL makes the new process more effective.
\item {} We have deployed the proposed method in a production environment, integrated it into multiple product lines, and compared it with mainstream NL2SQL methods. ChatBI has achieved optimal results in terms of accuracy and efficiency.
\end{itemize}

In the following sections, we will provide details of \textit{ChatBI}. 
In section 2, we present necessary backgrounds. 
Section 3 introduces the overview of \textit{ChatBI}. 
We then introduce the \textbf{Multi-Round Dialogues Matching} and \textbf{Single View Selection} in Section 4 and the \textbf{Phased Process Flow} in Section 5. 
We evaluate our method in Section 6. 
Finally, we conclude this paper in Section 7.

%% file: 2.background.tex
\begin{figure}[t]
  \includegraphics[width=0.495\textwidth]{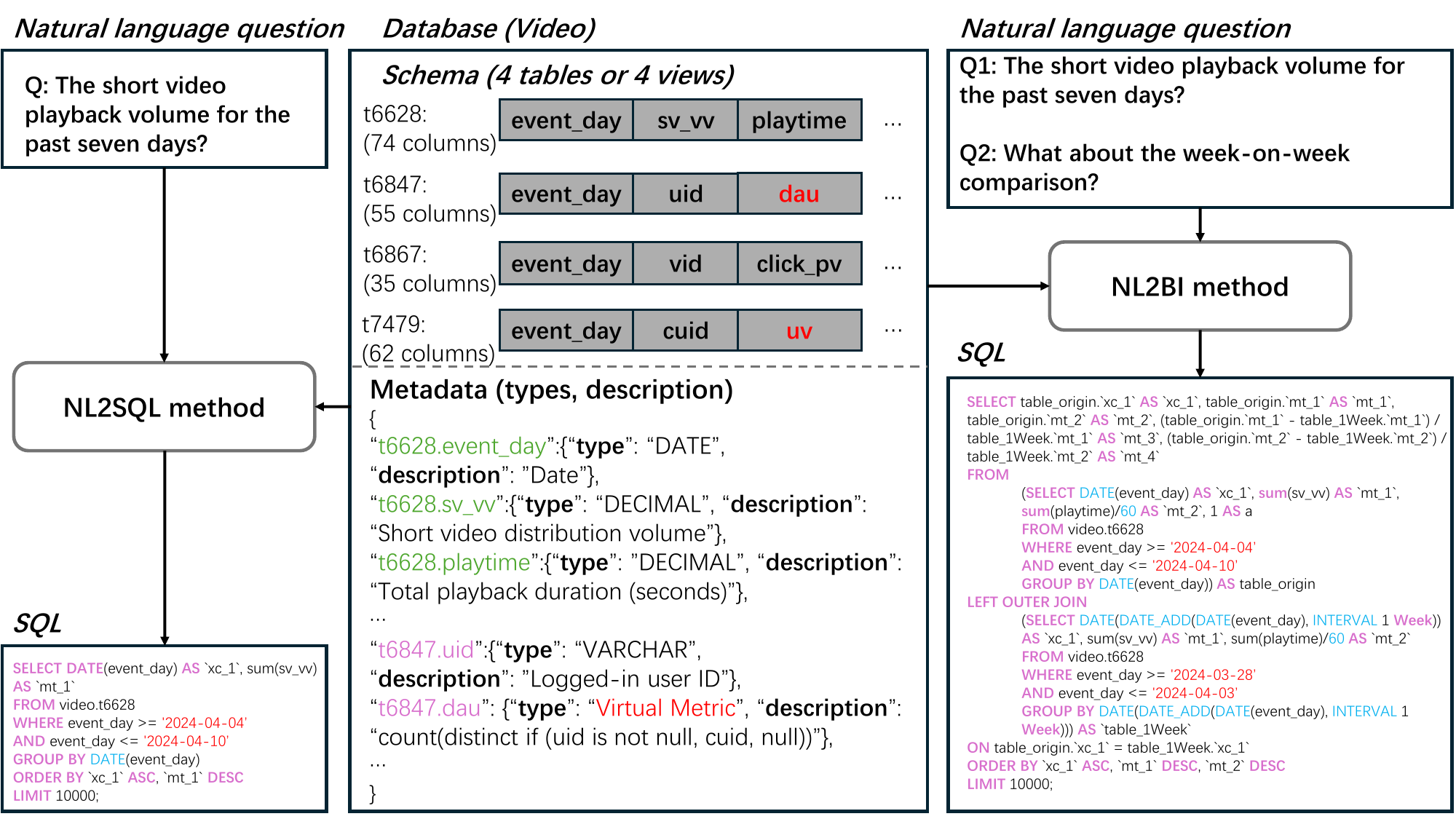}
  \caption{An example of metadata in NL2SQL and NL2BI. ``dau'' and ``uv'' are Virtual Columns.}
  \Description{The BI example.}
  \label{figure:SchemaExample}
\end{figure}

\section{PRELIMINARY}

\noindent
\textbf{NL2SQL Task. }
The primary goal of the NL2SQL task is to construct a SQL query \textit{S} from a given Natural Language question \textit{Q}, a database \textit{D} and a corresponding database table \textit{T}. 
The \textit{S} should be executable and accurately address the informational intent encapsulated in \textit{Q}, thereby producing a valid response within the context provided by \textit{D} and \textit{T}: 
\begin{equation}
    S\ =\ Parser(Q,D,T),
\end{equation}
where the Parser() is designed to analyze \textit{Q} leveraging \textit{D} and \textit{T}. 
\textit{D} contains database schema and database metadata, while \textit{T} contains database metadata which contains column types and descriptions. 
An example of the metadata is presented in Figure~\ref{figure:SchemaExample}.

\noindent
\textbf{NL2BI Task. }
The primary goal of the NL2BI task is to construct a SQL query \textit{S} from given Multi-Round Natural Language question \textit{MRQ}, a database \textit{D} and a corresponding database view \textit{V}: 
\begin{equation}
    S\ =\ Parser(MQ,D,V),
\end{equation}

\noindent
\textbf{Virtual Column. }
Virtual columns are columns that are typically calculated in BI scenarios and their corresponding rules, which do not physically exist in tables or views. 
As shown in Figure~\ref{figure:SchemaExample}, within table or view ``t6847'', there is a column named ``uid'', which represents the logged-in user ID. 
In the BI scenario, DAU is a crucial column, standing for Daily Active Users. 
It signifies the number of active users per day. 
In ``t6847'', ``dau'' exists as an virtual column.
The description associated with ``t6847.dau'' is ``\textit{count( distinct if(uid is not null, uid, null) )}'', which specifies the computation syntax for obtaining DAU.

%% file: 3.tech.tex
\section{System Overview}

\begin{figure*}[t]
  \includegraphics[width=0.98\textwidth]{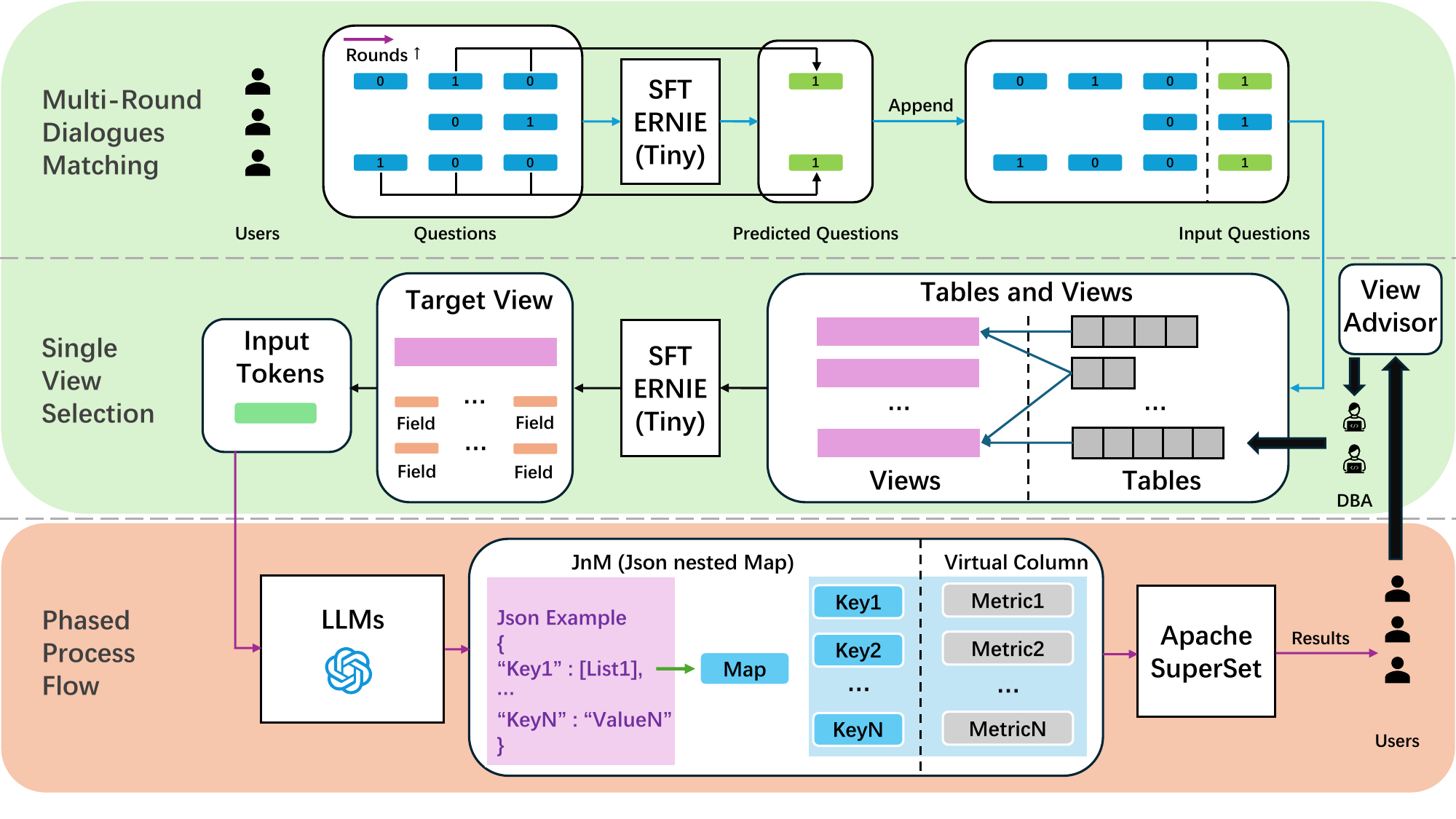}
  \caption{The Overview of ChatBI. }
  \Description{The overview of ChatBI.}
  \label{figure:Overview}
\end{figure*}

Figure~\ref{figure:Overview} presents an overview of ChatBI, which comprises three modules: \textbf{Multi-Round Dialogues Matching}, \textbf{Single View Selection}, and a \textbf{Phased Process Flow}. 
Specifically, after a user poses a Natural Language question, the Multi-Round Dialogues Matching module determines whether the question is a follow-up in a Multi-Round dialogue or a standalone query in a Single-Round dialogue to ensure the question encapsulates the user's complete intent. 
The Single View Selection module identifies the corresponding view that matches the user's intent based on the question containing the full intent, and organizes this corresponding view and the question into prompts that are passed to a LLM. 
Unlike previous process flows that directly generate the target SQL from these prompts, the phased prcess flow first requires the LLM to output a JSON intermediary. 
This intermediary serves as the input for a rule-based SQL generation method, producing the target SQL. 
Existing rule-based methods struggle with complex semantics, computations, and comparison relationships.
Therefore, ChatBI introduces a method of Virtual Column in the phased process flow. 
This method avoids the need for the LLM to learn complex semantics, computations, and comparison relationships, reducing the difficulty of the problem-solving process for the LLM in the phased process flow. 
Finally, with the help of Apache SuperSet~\cite{Superset}, ChatBI enables the output of custom results in various formats (e.g., charts, line graphs). 
After each query is executed, it is evaluated by users to determine if the generated SQL resolves the query. 
If resolved, it is counted in the \textit{View Advisor}. 
And if not, it undergoes analysis, and cases where the failure is due to the absence of a view, prompting the Single View Selection module to fail, are assessed by the DBA to decide if a new view should be established. 
This ensures the accuracy of the Single View Selection module. 
Based on the counts, the DBA can also dynamically delete seldom-used views, achieving dynamic management of views.

\section{Balances Economy and Accuracy.}

In this section, we first discuss Multi-Round Dialogue matching in Section~\ref{section:MRDMatching}, which addresses the interactivity of user queries. 
Subsequently, in section~\ref{section:singleViewSelection}, we introduce the Single View Selection method designed to meet the challenges posed by the large number of columns and column ambiguities in BI scenarios. 

\subsection{Multi-Round Dialogue Matching}\label{section:MRDMatching}

\begin{figure}[t]
  \includegraphics[width=0.485\textwidth]{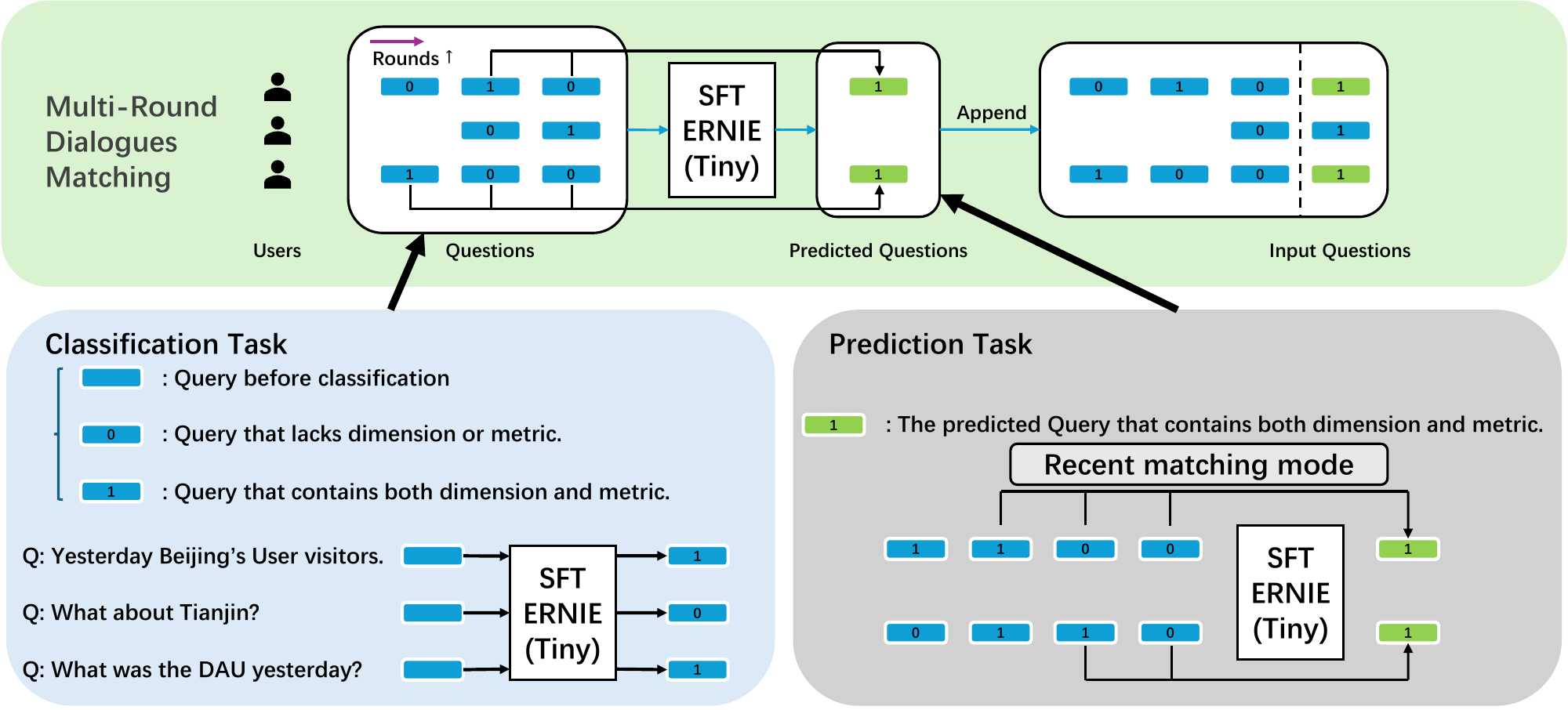}
  \caption{The method used in MRD Matching. }
  \Description{The overview of ChatBI.}
  \label{figure:MRDMatching}
\end{figure}

In BI scenarios, dimensions and columns are the two most crucial types of features. 
Dimensions typically correspond to statistical times or granularity, such as ``the past three days'' representing a time dimension, while columns correspond to specific columns, such as the ``User visitors'' column representing the number of user visits. 
Typically, when a query contains both dimensions and columns, such as the query ``User visits in the Beijing city yesterday'', it is possible to generate the corresponding SQL, and such a query can be considered as the current query. 
However, when a query lacks either a dimension or a column, for example, ``What about Tianjin?'', it cannot generate the corresponding SQL, thus it is necessary to treat it as part of a MRD query, using historical queries as sequence input to predict the current query. 
As shown in Figure~\ref{figure:MRDMatching}, ChatBI splits the MRD matching problem into two tasks: a classification task and a prediction task. 

\noindent
\textbf{Classification Task.}
From the perspective of the timeliness of calls in actual production systems, ChatBI uses a Pre-training model rather than a LLM to perform this text classification task. 
In text classification tasks, the State-of-the-art (SOTA) Pre-training method is Bert~\cite{yin2020tabert}. 
In our production system, where the user input is in Chinese, we use the ERNIE~\cite{sun2019ernie} model, which is a Bert-like model that performs better on Chinese NLP tasks. 
The effectiveness of this classification model is demonstrated by the following formula $I(Q)$: 
\begin{equation}
    I\left( Q \right) =\begin{cases}
	1&		if\ Q\ contains\ both\ dimensions\ and\ columns,\\
	0&		if\ Q\ lacks\ either\ dimensions\ or\ columns.\\
\end{cases} \\
\end{equation}
Next, we detail the specifics of the Pre-training dataset. 
We employ a straightforward method to construct the Pre-training corpus. 
Firstly, for Q labeled as 1, which includes queries containing both dimensions and columns, the distinction between whether a column in a database table or view is a dimension or column is quite clear. 
For instance, ``event$\_$day'' typically represents a time dimension, and ``app$\_$id'' often represents different business lines, also a dimension. 
Columns, such as ``sv$\_$vv'' which represents the distribution volume of short videos, and ``dau'' which represents daily active users, are clearly defined as such. 
Simply concatenating dimension and column can generate Q labeled as 1, such as ``event$\_$day + sv$\_$vv'' could produce the training data: ``What is the volume of short video plays over the past three days?''. 
For Q labeled as 0, it is natural to create data containing only one of dimensions or columns by removing either dimension or column from data labeled as 1. 
For data that contains neither dimensions nor columns, we utilize GPT-4 for generation. 
The training model employs a 4-layer transfer$\_$former ERNIE (Tiny) model, which first classifies each user query. 

\noindent
\textbf{Prediction Task.}
Starting from the perspective of timeliness, ChatBI does not directly use LLMs for prediction. 
Queries labeled as 1 by the classification model can be directly treated as current queries, while those labeled as 0 require processing to become current queries. 
An initial idea is to use all of a user's historical queries as sequence input to predict queries that meet the criteria of containing both dimensions and columns. 
This approach has been found to be ineffective (details discussed in Section~\ref{section:experiment}). 
We observe that user queries mostly have logical coherence and often follow up on previous queries, so the most recent queries labeled as 0 are typically supplements or follow-ups to the most recent queries labeled as 1. 
Based on this observation, we have designed a \textit{recent matching mode}: for a query, we identify the timestamp of the most recent query classified as 1 in the user’s historical inputs, and from this timestamp to the current moment, we take all queries posed by the user as a sequence to a 4-layer transfer$\_$former ERNIE (Tiny) for prediction, treating the predicted query as the current query. 

\begin{algorithm}[t]
  \caption{MRDs Matching Algorithm }
  \label{alg:MRDs}
  \begin{algorithmic}[1]
    \Require
      MRDs queries $\{Q_{1}, Q_{2}, ... , Q_{N}\}$ ($Q_{N}$ is the latest query) 
    \Ensure A query $Q$ is labeled as one
    
    \If {$I(Q_{N})$ is Zero}
        \State {$ID$ $\leftarrow$ -1}
        \For{$i \leftarrow N, N-1,  \ldots,  2, 1$} \label{line:check1}
            \If {$I(Q_{i})$ is One}
                \State {$ID$ $\leftarrow$ i}
                \State \textbf{break}
            \EndIf
        \EndFor \label{line:check1.1}
    \Else
        \State{\textbf{Return} $Q_{N}$} \label{line:check2}
    \EndIf

    \If {$ID$ $\neq$ -1}

        \State{Use $\{Q_{ID}, Q_{ID+1}, \ldots, Q_{N}\}$ as the inputs, and use the SFT ERNIE Model to predict the $Q_{N+1}$} \label{line:check3}
        
        \State{\textbf{Append} $Q_{N+1}$ to MRDs queries} \label{line:check4}
    \Else
        \State{\textbf{Return} \textbf{None}}
    \EndIf

    \State {\textbf{Return} $Q_{N+1}$}
    \end{algorithmic}
\end{algorithm}

In summary, Algorithm~\ref{alg:MRDs} illustrates the main procedures of the MRDs matching algorithm. 
For the user's MRDs, when the latest dialogue contains both \textit{dimensions} and \textit{columns}, it can be directly used as the current query input (Line~\ref{line:check2}). 
Otherwise, we select the nearest dialogue with label one as the starting point and the latest dialogue as the endpoint (Line~\ref{line:check1} - Line~\ref{line:check1.1}), input all dialogues in the interval into the \textit{ERNIE} model to predict dialogues with label one (Line~\ref{line:check3}), and directly add them to the MRDs (Line~\ref{line:check4}). 
The user's current query is used as the Natural Language input required by the single view selection.

\subsection{Single View Selection}\label{section:singleViewSelection}

Schema linking is the process of selecting the target columns required by a Natural Language from all the columns in the data tables. 
This task presents a heightened challenge in BI scenarios due to the large number of columns, in contrast to ordinary scenarios. 
In previous work~\cite{floratou2024nl2sql}, LLMs were limited by the number of tokens, preventing them from processing tables with large number of columns, and the longer the prompt, the more severe the degradation of the model's capabilities. 
Additionally, this work~\cite{floratou2024nl2sql} mentions that the optimal method for schema linking is to provide only the optimal tables and columns (29.7\%), followed by providing only the optimal tables (24.3\%), and then the LLM's own selection (21.1\%). 
Advanced LLMs demonstrate superior performance on NL2SQL tasks~\cite{floratou2024nl2sql}, albeit with higher costs (GPT-4-32K is about 60X more expensive than GPT-3.5-turbo-instruct~\cite{oepnAIPricing} at the time of this writing). 
For DIN-SQL~\cite{pourreza2024din}, generating a single SQL query costs \$0.5, which is unacceptable in actual production systems. 
Therefore, both from the perspective of reducing the number of tokens and solving the schema linking problem, using a smaller and cheaper model to provide the optimal tables and columns has become the optimal choice. 

Meanwhile, it is noteworthy that \textit{Views}, despite their mature development in \textbf{data warehouses}, \textbf{data mining}, \textbf{data visualization}, \textbf{decision support systems}, and \textbf{BI systems}, are absent in the \textbf{Spider}~\cite{yu2018spider} and \textbf{BIRD}~\cite{li2024can} datasets.
These \textit{Views} are established upon business access and exist objectively within real data tables. 
The presence of \textit{Views} substantially mitigates the need for \textbf{JOIN} operations between tables, enabling the execution of queries that would otherwise be dispersed across multiple tables on a single table. 
And if there exists an optimal single view that contains the optimal columns, the view is the answer. 
Otherwise, there may be two or more views that require a join operation to be resolved or other errors. 
We report all such join scenarios to the DBA in \textit{View Adviosr}, who evaluates the proportion of these types of queries and decides whether to establish views for such queries to avoid join operations. 
Consequently, in the context of NL2BI tasks, an optimal strategy for decomposing the problem is to convert the schema linking issue into a \textbf{Single View Selection problem}. 

Given multiple Views \textit{Vs}, each containing numerous columns that genuinely exist in the data tables without duplication, and each column potentially appearing in zero or more \textit{Vs}, the objective is to select a View \textit{V} that encompasses all the target columns \textit{Cols} corresponding to the user's Natural Language query while minimizing the total number of columns in the \textit{V}. 
The \textbf{Single View Selection problem} is defined as follows: 
\begin{equation}
    V\ =\ Select(Vs, Cols),
\end{equation}
where the Select() is designed to find a \textit{V} that contains \textit{Cols} in \textit{Vs}. 

The abundance of \textit{Cols} in BI scenarios poses a challenge when inputting all schemas as tokens into LLMs, leading to the issue of token number exceeding the limit. 
Furthermore, the same column across different tables in the same business may be expressed in multiple ways, resulting in column ambiguity. 
Addressing the \textbf{Single View Selection problem} can significantly compress the token count of the schema linking results, enabling the LLMs to accommodate schemas. 
The logical matching relationships established during \textit{V} creation effectively resolve column ambiguity. 

\begin{figure}[t]
  \includegraphics[width=0.495\textwidth]{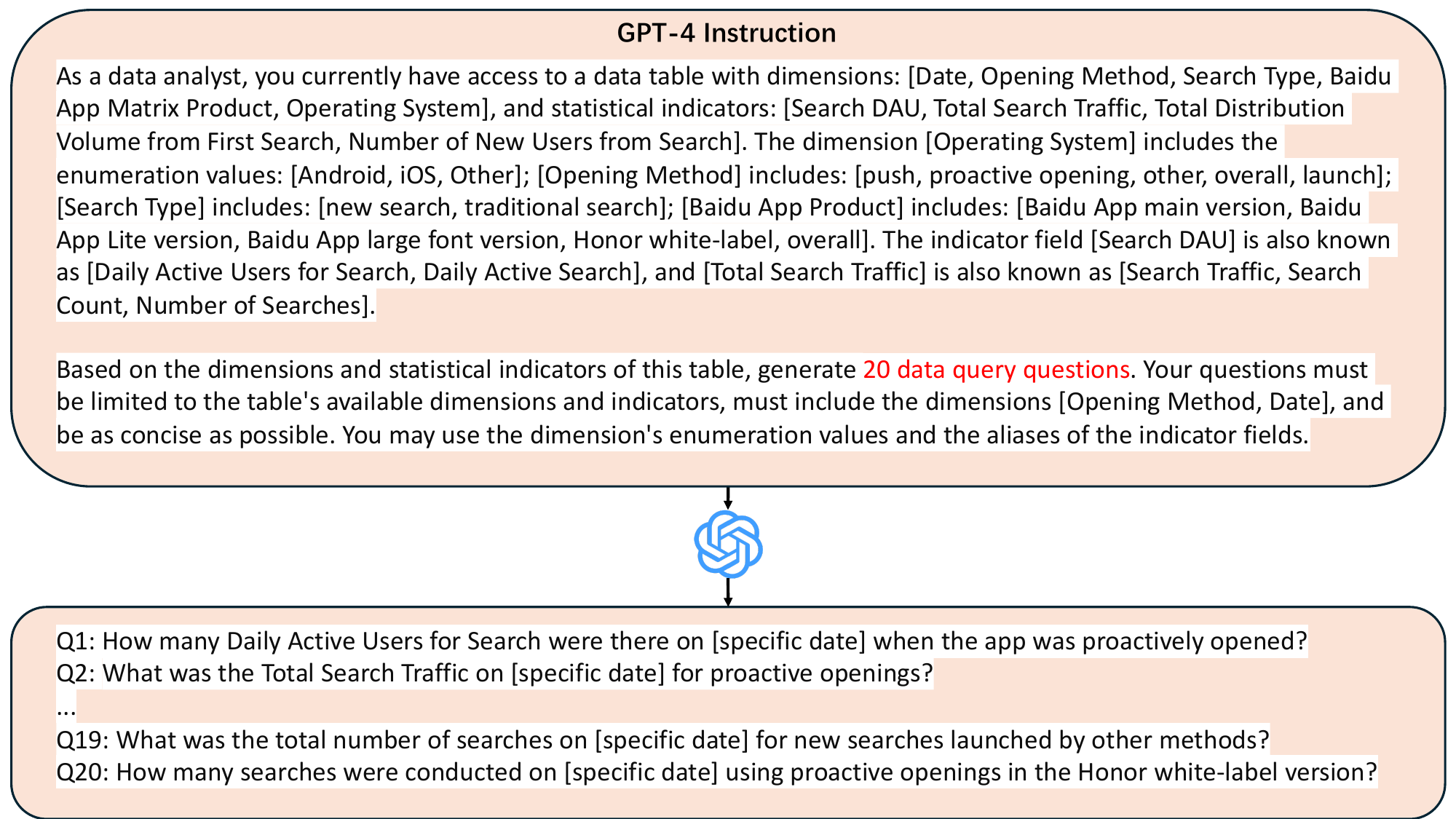}
  \caption{The prompt for SFT training data generation using GPT-4. }
  \Description{Single View Selection Prompt.}
  \label{figure:SingleViewSelectionPrompt}
\end{figure}

\begin{table}[t]
\caption{The accuracy of each view selection model.}
\label{exp:acc}
\begin{tabular}{c|c}
\hline
Baseline                 & ACC    \\ \hline
SFT ERNIE (4 layers)     & 95.7\% \\ \hline
Bert (12 layers)         & 95.2\% \\ \hline
CNN (1 layer, 3-channel) & 83.9\% \\ \hline
\end{tabular}
\end{table}

Considering online invocation and economic benefits, we decide to first adopt smaller and cheaper models and assess whether they can effectively complete view selection. 
In line with the initial approach of not employing a LLM to tackle MRD scenarios, we designed and compared three lightweight models for practical application: a Convolutional Neural Network (\textit{CNN}) model (1 layer and 3-channel), a Bert model (12 layers) and an \textit{SFT ERNIE} model (4 layers) for view selection data. 
The training data, generated as shown in Figure~\ref{figure:SingleViewSelectionPrompt}, specifically comprises real \textit{Views} and their included \textit{Cols}, a capability inherent to BI scenarios. 
We use this method to generate over 3,000 entries as training data and 500 entries as test data. We then validate three models with the results presented in Table~\ref{exp:acc}.
Based on the comparison of experimental results, the \textit{SFT ERNIE Tiny} model was ultimately selected. 

ChatBI also maintains a \textit{View Advisor} module within its Single View Selection. 
After ChatBI finishes running, all queries that fail the single view selection are collected into the View Advisor and handed over to the business line's DBA for evaluation. 
The DBA assesses the proportion of such queries in user scenarios and decides whether to establish views to support such queries and avoid join operations.
Meanwhile, all queries that successfully select a single view increment the hit count for the corresponding view by one. 
In practice, through the View Advisor, the DBA can easily add new views and delete views that have not been used for a long time.

\section{Decompose Problems before Solving them.}\label{sec:post}

\begin{figure}[t]
  \includegraphics[width=0.495\textwidth]{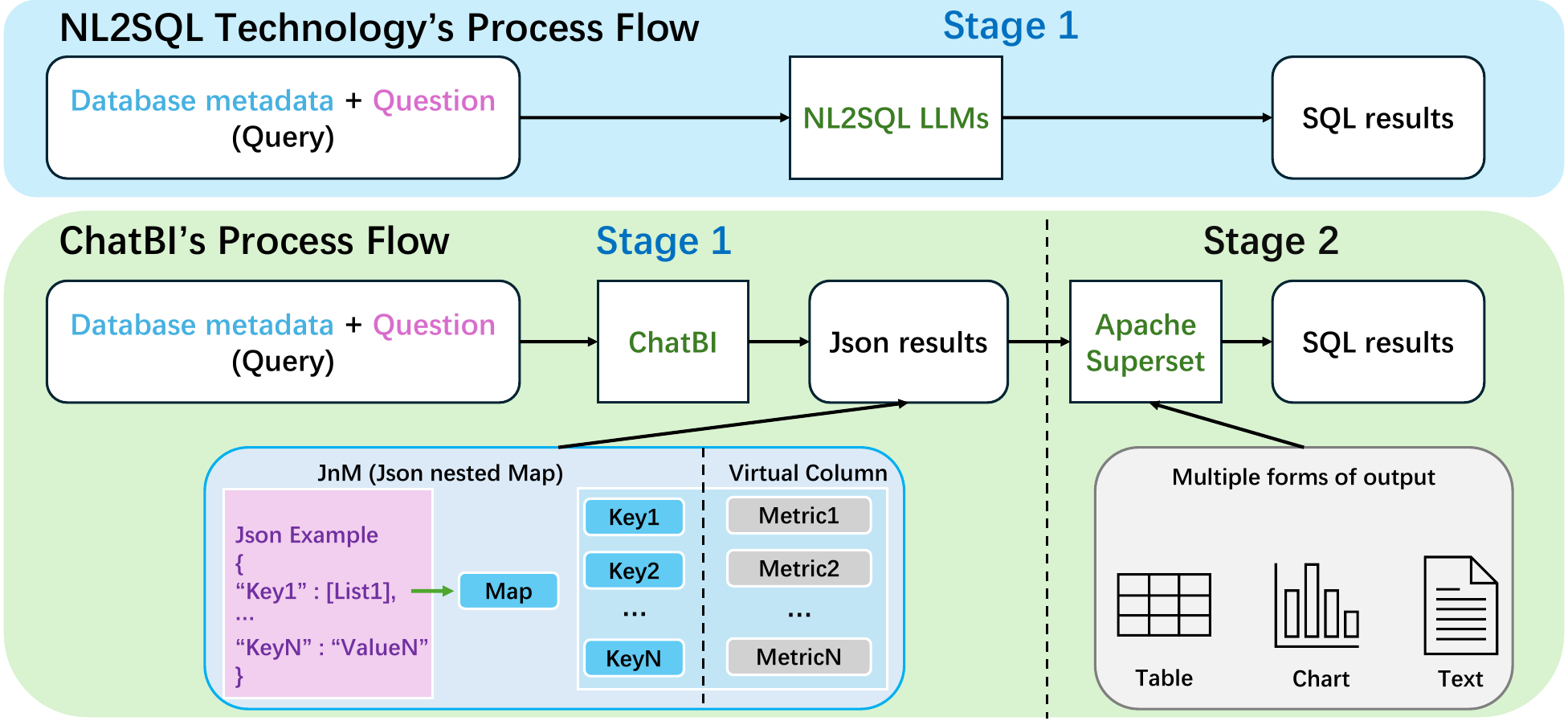}
  \caption{The difference between ChatBI and other NL2SQL technology's process flow. }
  \Description{The decomposed problem.}
  \label{figure:decomposedProblem}
\end{figure}

In this section, we first discuss a phased process that differs from previous methods in Section~\ref{section:phasedProcess}. 
Subsequently, in Section~\ref{section:VirtualColumn}, we explore how ChatBI employs \textit{Virtual Columns} within this process. 

\subsection{Phased Process Flow}\label{section:phasedProcess}

In BI scenarios, complex semantics, computational relationships, and comparison relationships make it difficult for existing methods to handle NL2BI tasks effectively. 
Current process flows rely solely on SFT data or prompts, and it is challenging to enumerate all relationships and provide their corresponding golden SQL. 
Even if these relationships and their golden SQL are obtained, incorporating them into prompts is not feasible, as it would lead to token overflow and, concurrently, the trimming could impair the LLM's understanding of other knowledge. 
Recent studies have demonstrated that the performance of LLMs on complex tasks can be enhanced through a decomposed prompting technique~\cite{khot2022decomposed}. 
This method entails segmenting a task into multiple steps and employing the intermediate results to synthesize a final answer. 
Unlike algebraic expressions, which consist of clearly delineated steps or operations, the deconstruction of a complex SQL query poses a significantly greater challenge due to the declarative nature of SQL and the intricate interrelations among various query clauses~\cite{pourreza2024din}. 
In response to these challenges, ChatBI introduces a phased process flow designed to decompose the NL2BI problem, specifically aiming to effectively handle complex semantics, computational relationships, and comparison relationships within BI scenarios. 

As shown in Figure~\ref{figure:decomposedProblem}, ChatBI introduces a phased process flow that initially utilizes a LLM to generate JSON-formatted intermediate outputs. 
These outputs are then fed into BI middleware such as Apache SuperSet, which supports various output formats, to display results. 
The key difference in the new process flow compared to the traditional reliance on LLMs for SQL generation is its reliance solely on generating JSON. 
This approach does not require, as DAIL-SQL~\cite{gao2023text} does, providing extensive Natural Language and its corresponding golden SQL in the prompt, which significantly reduces the number of tokens. 

The new process flow does not require the LLM to understand the complex semantics associated with various SQL join operations, nor does it need to grasp specific comparison relationships like DtD (Day-to-Day) comparisons typical in BI scenarios, or the complex computational relationships among columns. 
By decoupling the complexity of SQL problems from the LLM, the difficulty of generating JSON in the new process is merely Simple (according to the task grading by DIN-SQL and MAC-SQL based on the complexity of SQL generation tasks, since ChatBI does not need to understand SQL, all are classified as Simple tasks). 
This significantly enhances the accuracy of the process flow. 
The step of generating SQL is executed within Apache SuperSet, where extensive research from the database community on generating SQL based on dimensions and columns already exists, such as CatSQL~\cite{fu2023catsql}, which combines deep learning models with rule-based models. 
ChatBI implements multiple universal templates in Apache SuperSet. 
And the JSON output from the LLM acts like filling in the blanks for the corresponding query template, eventually outputting SQL. 

\subsection{Virtual Column}\label{section:VirtualColumn}

Although the phased process flow, which first breaks down NL2SQL into NL2JSON, reduces the task's complexity, the lack of SQL understanding cannot be simply compensated by templates in Apache SuperSet. 
Particularly for columns like ``DAU'', which require computational relationships derived from other columns, the complexity of these relationships significantly reduces the accuracy of the templates. 
Therefore, ChatBI introduces the concept of \textit{Virtual Columns}, as illustrated in Figure~\ref{figure:SchemaExample}. 
For the ``dau'' column, it does not physically exist in the data table or view but merely stores the SQL computation relationships needed to derive it. 
\textit{Virtual columns} like these can be accessed through their corresponding key (column name) to find the value (computation rule). 
We represent virtual columns using a map structure nested into the JSON, referred to as \textit{JnM}, ensuring that all queries hitting Virtual Columns directly access the stored computation rules.

Although the computational rules for columns like ``DAU'' in the BI column are relatively standard, the NL2BI task possesses its uniqueness. 
However, we believe that when applying LLMs to specific tasks, they can be broken down into NL2JSON like ChatBI, utilizing a map stored in the JSON to manage complex rules. 
This approach not only reduces the difficulty of generating tasks for LLMs, enhancing their accuracy, but also cleverly uses JnM to perceive complex semantics and relationships. 
Relationships similar to those in virtual columns can be generated by LLMs, thereby facilitating caching to speed up computations in future calls.

In addition to ``DAU'', here are two examples of \textit{Virtual Columns} in ChatBI. 
``\textit{staytime}'' represents the video playback duration in seconds. 
A frequently asked column is the total playback duration in minutes, whose \textit{Virtual Column}, \textit{stay\_time\_min}, is stored in the view's schema with the calculation rule ``sum(staytime) / 60''. 
Another commonly queried column, the average playback duration per user, \textit{stay\_time\_per}, is also stored in the view's schema, with the calculation rule ``sum(staytime)/60/count(uid)'', where ``\textit{uid}'' represents the user ID.

%% file: 5.expr.tex
\section{Experimental Evaluation}\label{section:experiment}

\begin{figure*}[t]
  \includegraphics[width=0.98\textwidth]{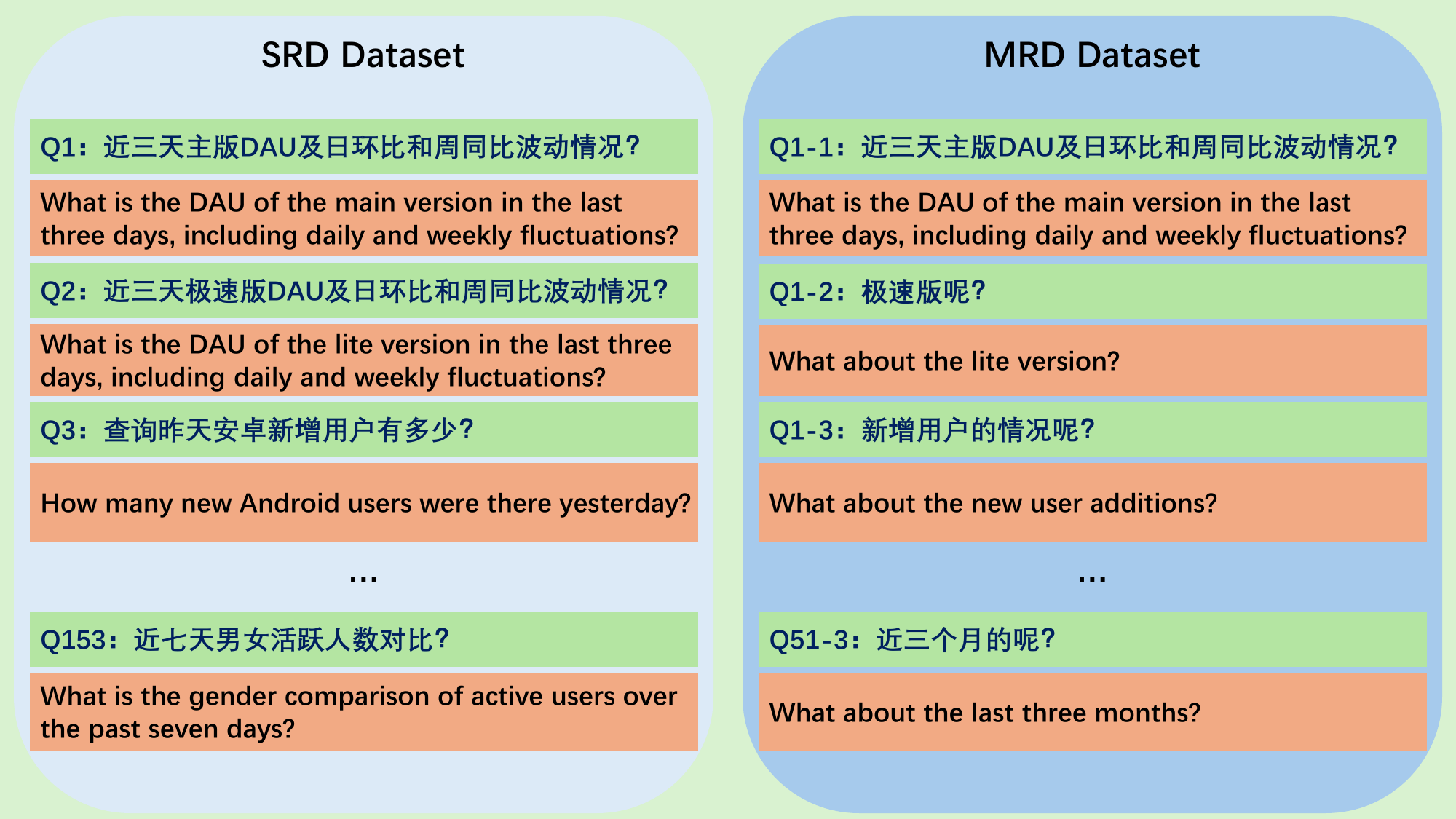}
  \caption{Introduction to the SRD dataset and MRD dataset. The main version and the lite version correspond to different apps. DAU stands for Daily Active Users, and new users refer to users who are registering for the first time. }
  \Description{The SRD Dataset and MRD Dataset.}
  \label{figure:Dataset}
\end{figure*}

\subsection{Experimental Setup}

\begin{table}[t]
\caption{LLM leaderborad~\cite{LLMRanking} 2024-03. Avg means the Average performance, Lg means the Language and Knowl means the Knowledge}
\label{exp:leaderBorad}
\begin{tabular}{ccccccc}
\hline
LLM           & O/C Source   & Avg & Lg & Knowl & Agent \\ \hline
GPT-4-Turbo   & Close Source & 62      & 54.9     & 66.3            & 82    \\
Claude3-Opus  & Close Source & 60.5    & 44.6     & 68.9            & 81.1  \\
GLM-4         & Close Source & 57.8    & 57.7     & 71                & 73.4  \\
Qwen-MAX-0107 & Close Source & 55.8    & 59.4     & 55.4            & 60    \\
Qwen-MAX-0403 & Close Source & 55.6    & 58.8     & 70.4            & 73.7  \\
Qwen-72B      & Open Source  & 54.5    & 59.6     & 67.8            & 54.6  \\ \hline
\end{tabular}
\end{table}

\noindent
\textbf{Datasets}
The Spider~\cite{yu2018spider} and BIRD~\cite{li2024can} datasets are commonly utilized to evaluate the efficacy of NL2SQL parsers across diverse databases. 
The two datasets are used in the NL2BI evaluation context, where they adopt a SRD mode and overlook the significant number of existing views. 
The presence of views can optimize most cross-table queries within a single view. 
Therefore, we have developed two datasets from real BI scenario data, \textbf{SRD dataset} and \textbf{MRD dataset} without \textbf{golden SQLs}, each provided with 4 views, as shown in Figure~\ref{figure:Dataset}. 
Each dataset contains 226 columns and 4 views and encompasses 153 real BI Chinese queries. 
To enhance the difficulty of queries in the \textbf{MRD dataset}, 153 queries consist of 51 three-round dialogues. 
Among these dialogues, 19 involve queries that switch views, meaning the tables accessed in the three rounds of dialogue differ. 
We consider these 19 dialogues to represent challenging tasks. 
BIRD and Spider provide a golden SQL for each query, a process that is quite challenging to construct. 
Therefore, from the perspective of actual production systems, both \textbf{SRD dataset} and \textbf{MRD dataset} do not have golden SQLs.


\noindent
\textbf{Evaluation Metrics}
Additionally, these NL2SQL technologies employ \textit{exact string match} and \textit{execution match} metrics. 
The former only compares whether results share the same string expression, which may lead to many correct results being considered incorrect. 
Relying solely on execution results depends on the dataset's predefined golden SQL, which could also lead to misjudgments. 
For instance, in a BI context, for the query: ``\textit{Yesterday's interaction metrics for users of app1}'', the following golden SQL is as follows:
\begin{lstlisting}[ language=SQL,
                    deletekeywords={IDENTITY},
                    deletekeywords={[2]INT},
                    morekeywords={clustered},
                    framesep=8pt,
                    xleftmargin=40pt,
                    framexleftmargin=40pt,
                    frame=tb,
                    framerule=0pt ]
SELECT SUM(click_pv) AS total_click_pv, 
     SUM(disp_pv) AS total_disp_pv, 
     SUM(play_pv) AS total_play_pv, 
     SUM(like_pv) AS total_like_pv, 
     SUM(comment_pv) AS total_comment_pv, 
     SUM(collection_pv) AS total_collection_pv, 
     SUM(shareto_pv) AS total_shareto_pv, 
     SUM(follow_pv) AS total_follow_pv
FROM table1
WHERE event_day = DATE_SUB(CURRENT_DATE, INTERVAL 1 DAY) 
AND appid = 'app1';
\end{lstlisting}
These columns click\_pv, disp\_pv, play\_pv, like\_pv, comment\_pv, collection\_pv, shareto\_pv, and follow\_pv are interaction metrics. 
Yet any non-empty set containing one or several of click\_pv, disp\_pv, play\_pv, like\_pv, comment\_pv, collection\_pv, shareto\_pv, or follow\_pv can be considered correct. 
The SQL below is also deemed correct:
\begin{lstlisting}[ language=SQL,
                    deletekeywords={IDENTITY},
                    deletekeywords={[2]INT},
                    morekeywords={clustered},
                    framesep=8pt,
                    xleftmargin=40pt,
                    framexleftmargin=40pt,
                    frame=tb,
                    framerule=0pt ]
SELECT SUM(click_pv) AS total_click_pv, 
     SUM(disp_pv) AS total_disp_pv
FROM table1
WHERE event_day = DATE_SUB(CURRENT_DATE, INTERVAL 1 DAY) 
AND appid = 'app1';
\end{lstlisting}
Following Previous work~\cite{floratou2024nl2sql}, we consider using the metric \textit{usefulness} rather than \textit{correctness}.
So we use the metric useful execution accuracy (UEX) which compares the execution output of the predicted SQL query and the intent of this query. 
SQL queries with column errors and syntax errors are invariably considered incorrect. 
The results of all executable queries are assessed by experienced DBAs to determine whether they satisfy the user's intent for that specific query. 
We also provide the number of tokens used in the prompts and responses for each method to assess a certain economic cost. 
We use prompt tokens (PT) to represent the number of tokens used in the prompts, and response tokens (RT) to represent the number of tokens used in the responses. 

\noindent
\textbf{Baselines}

\begin{itemize}
\item {\textbf{DIN-SQL}~\cite{pourreza2024din}.} Din-SQL deconstructs the NL2SQL task into discrete subtasks, formulating distinct prompts for each subtask to guide GPT-4 through the completion of each component, thereby deriving the final SQL.
\item {\textbf{MAC-SQL}~\cite{wang2023mac}.} MAC-SQL utilizes multi-agent collaboration for NL2SQL tasks and reach an accuracy of 59.6\% BIRD. 
\item {\textbf{ChatBI}.} ChatBI employs all the modules introduced in this paper, matching Multi-Round Dialogues and performing Single View Selection through a phased processing flow. Initially, a LLM outputs JSON, which is subsequently used to generate SQL that predicts the SQL corresponding to the query. 
\item {\textbf{ChatBI w/o PF + zero-shot}.} The primary difference between ChatBI w/o PF + zero-shot and ChatBI is that it does not employ a phased process flow. Similar to other NL2SQL methods, it directly generates the corresponding SQL using a LLM. 
\end{itemize}

\subsection{Evaluations on SRD Dataset}

\begin{table}[t]
\caption{Useful execution accuracy (UEX), prompt tokens (PT) and response tokens (RT) on SRD dataset.}
\label{exp:SRDDataset}
\begin{tabular}{c|c|c|c}
\hline
Baseline      & UEX     & PT      & RT    \\ \hline
DIN-SQL       & 20.92\% & 1472242 & 53332 \\ \hline
MAC-SQL       & 52.29\% & 1272661 & 36952 \\ \hline
ChatBI        & 74.43\% & 841550  & 29336 \\ \hline
ChatBI w/o PF + zero-shot & 58.17\% & 586640 & 28886 \\ \hline
\end{tabular}
\end{table}

It is important to note that the all experiments utilize the ``2023-03-15-preview'' version of GPT-4~\cite{openAIGPT} according to Table~\ref{exp:leaderBorad} and ChatBI uses the Erniebot-4.0~\cite{ErineBot} (The work is also done by Baidu Inc.) in actual production systems due to data privacy concerns.  
In Table~\ref{exp:SRDDataset}, we report the performance of our method and baseline methods on the SRD dataset. 
It is evident our method surpasses other baseline methods using less tokens on both prompt and response. 

\begin{table}[t]
\caption{Useful execution accuracy (UEX), prompt tokens (PT) and response tokens (RT) on SRD dataset of ChatBI w/o PF + zero-shot and ChatBI w/o PF + few-shot.}
\label{exp:DifferentShots}
\begin{tabular}{c|c|c|c}
\hline
Baseline                  & UEX     & PT & RT \\ \hline
ChatBI w/o PF + zero-shot & 58.17\% & 586640 & 28886 \\ \hline
ChatBI w/o PF + few-shot  & 49.67\% & 735144 & 24126 \\ \hline
\end{tabular}
\end{table}

\noindent
\textbf{ChatBI achieves higher UEX with fewer tokens on SRD dataset.} 
Compared to MAC-SQL, ChatBI improves the UEX metric by \textbf{42.34\%} ($\frac{(74.43-52.29)}{52.29}$), using \textbf{33.87\%} ($\frac{(1272661-841550)}{1272661}$) fewer tokens in the prompt stage and \textbf{20.61\%} ($\frac{(36952-29336)}{36952}$) fewer tokens in the response stage. 
Relative to DIN-SQL, ChatBI increases the UEX metric by \textbf{255.78\%} ($\frac{(74.43-20.92)}{20.92}$), utilizing \textbf{42.84\%} ($\frac{(1472242-841550)}{1472242}$) fewer tokens in the prompt and \textbf{44.99\%} ($\frac{(53332-29336)}{53332}$) fewer tokens in the response. 
ChatBI and MAC-SQL both provide column values in their prompts, which offers significant adaptability for BI scenarios that require explicit column specifications, and it largely prevents SQL execution errors caused by the hallucinations of LLMs. 
In contrast, DIN-SQL does not provide column values in its prompts, resulting in a higher likelihood of errors in the filter conditions of the generated SQL. 
Furthermore, during the execution of DIN-SQL, it has been observed that it fails to analyze queries in Chinese. 
This issue arises because its prompts consist entirely of English without any Chinese elements. 
In cases where DIN-SQL produces unanalyzable SQL for Chinese queries, we re-execute it until it outputs SQL. 
Therefore, we believe that ChatBI achieves higher accuracy with fewer tokens used in the NL2BI SRD task.

\noindent
\textbf{The phased process flow assists LLMs in understanding complex semantics, computations, and comparison relationships. }
Although ChatBI uses \textbf{43.45\%} ($\frac{(841550-586640)}{586640}$) more tokens in prompts compared to ChatBI w/o PF + zero-shot, it achieves a \textbf{27.95\%} ($\frac{(74.43-58.17)}{58.17}$) improvement in UEX. 
We find that this is primarily because the phased process flow enables the LLM to bypass the need to comprehend complex semantics, computations, and comparison relationships directly within SQL. 
Instead, it only needs to understand how to map these complex relationships to the outputs required in JSON. 
By reducing the complexity of the task, the LLM can provide more accurate results, leading to an enhancement in UEX. 
Although the phased processing flow does not require the LLM to perceive SQL directly, it still necessitates an understanding of the JSON data type and the various parameter forms needed for subsequent SQL generation, hence more tokens are used in the prompts.

\noindent
\textbf{Longer prompts do not necessarily lead to better results. }
ChatBI w/o PF + zero-shot, which merely changes its output format from JSON to SQL compared to ChatBI, prompted us to explore whether increasing the number of shots could enable direct SQL generation to achieve or surpass the performance of JSON generation. 
In Table~\ref{exp:DifferentShots}, we examine the UEX, PT, and RT for ChatBI w/o PF + zero-shot and ChatBI w/o PF + few-shot on the SRD dataset. 
The table shows that with more shots provided, ChatBI w/o PF + few-shot actually experienced a \textbf{14.61\%} ($\frac{(58.17-49.67)}{58.17}$) decrease in UEX compared to ChatBI w/o PF + zero-shot, which uses \textbf{25.31\%} ($\frac{(735144-586640)}{586640}$) more tokens in prompts.  
Similarly, previous studies~\cite{floratou2024nl2sql} have also mentioned that longer prompts might cause degradation in large language models. 
Furthermore, shots that yield positive outcomes often require meticulous design~\cite{gao2023text}. 
Therefore, we do not extensively discuss the impact of shots in this paper.

\subsection{Evaluations on MRD Dataset}

\begin{table}[t]
\caption{Useful execution accuracy (UEX), prompt tokens (PT) and response tokens (RT) on MRD dataset.}
\label{exp:SRDDataset}
\begin{tabular}{c|c|c|c}
\hline
Baseline      & UEX     & PT      & RT    \\ \hline
DIN-SQL       & 20.26\% & 1526638 & 65850 \\ \hline
MAC-SQL       & 35.29\% & 1685080 & 30095 \\ \hline
ChatBI        &  67.32\% & 799442 & 20766 \\ \hline
ChatBI w/o PF + zero-shot  &  52.94\% & 592074 & 18723 \\ \hline
\end{tabular}
\end{table}

Since DIN-SQL~\cite{pourreza2024din} and MAC-SQL~\cite{wang2023mac} are not designed for Multi-Round Dialogues, we have made the following modifications in the prompts. 
The MRD Dataset consists of 51 sets of three-round queries, namely Query1, Query2, and Query3. 
The prompt for Query1 is as follows:
\begin{lstlisting} [language = c++]
/*
As a data analysis expert, you are to extract the necessary information from the data provided and output the corresponding SQL query based on the user's question. 
Let us consider the problem step by step. 
The current question is {Query1}.
*/
\end{lstlisting}
, the prompt for Query2 is shown below:
\begin{lstlisting} [language = c++]
/*
As a data analysis expert, you are to extract the necessary information from the data provided and output the corresponding SQL query based on the user's question. 
Let us consider these problems step by step. 
The last question is {Query1}, and the current question is {Query2}.
*/
\end{lstlisting}
, and the prompt for Query3 is also provided as follows:
\begin{lstlisting} [language = c++]
/*
As a data analysis expert, you are to extract the necessary information from the data provided and output the corresponding SQL query based on the user's question. 
Let us consider these problems step by step. 
The last two question are {Query1} and {Query2}, and the current question is {Query3}.
*/
\end{lstlisting}
.
In reality, within actual scenarios, LLMs do not know whether a query belongs to a MRD or a SRD for a better interactive experience. 
Unless it is explicitly required that users inform the LLM that the issue should be considered in conjunction with the previous question. 
In the experiments, ChatBI is not informed about which turn of a MRD the dialogue belongs to.
And this is entirely determined by the Multi-Round Dialogue Matching module.

\noindent
\textbf{ChatBI achieves higher UEX with fewer tokens on MRD dataset.} 
Compared to MAC-SQL, ChatBI improves the UEX metric by \textbf{90.76\%} ($\frac{(67.32-35.29)}{35.29}$), using \textbf{52.56\%} ($\frac{(1685080-799442)}{1685080}$) fewer tokens in the prompt stage and \textbf{31.00\%} ($\frac{(30095-20766)}{30095}$) fewer tokens in the response stage. 
Relative to DIN-SQL, ChatBI increases the UEX metric by \textbf{232.28\%} ($\frac{(67.32-20.26)}{20.26}$), utilizing \textbf{47.63\%} ($\frac{(1526638-799442)}{1526638}$) fewer tokens in the prompt and \textbf{74.09\%} ($\frac{(65850-20766)}{65850}$) fewer tokens in the response. 
As demonstrated in the evaluation of the SRD dataset, the presence of field hallucinations still results in DIN-SQL's SQL prediction accuracy being inferior to that of MAC-SQL and ChatBI. 
And ChatBI continues to achieve a higher UEX with fewer tokens in the MRD dataset. 

\noindent
\textbf{The Multi-Round Dialogue Matching module is better equipped to handle MRD scenarios.}  
The most significant change in the MRD dataset compared to the SRD dataset is that the questions are part of MRD, which introduces follow-up queries. 
For ChatBI, its UEX metric in the MRD dataset decreased by \textbf{9.55\%} ($\frac{(67.32-35.29)}{35.29}$) compared to the SRD dataset, while MAC-SQL decreased by \textbf{32.51\%} ($\frac{(52.29-35.29)}{52.29}$). 
This indicates that, compared to using LLMs to directly handle Multi-Round Dialogues, ChatBI's proposed Multi-Round Dialogue Matching module is more effective in managing MRD scenarios.

\noindent
\textbf{The Phased Process Flow and Virtual Columns are more adept at handling complex semantics, computations, and comparative relationships.} 
In the SRD dataset, it is observed that ChatBI, compared to ChatBI w/o PF + zero-shot, shows a \textbf{27.95\%} ($\frac{(74.43-58.17)}{58.17}$) improvement in UEX. 
Similarly, in the MRD dataset, ChatBI shows a \textbf{27.16\%} ($\frac{(67.32-52.94)}{52.94}$) increase in UEX compared to ChatBI w/o PF + zero-shot. 
We find that this is primarily because ChatBI achieves higher accuracy on SQL tasks involving complex semantics, computations, and comparisons in both datasets. 
Therefore, the Phased Process Flow and Virtual Column prove to be more effective in managing these challenging relationships.


\subsection{Discussion}

\noindent
\textbf{The limitations of the datasets primarily stem from the evaluation phase.} 
As introduced, both the SRD dataset and MRD dataset contain only 153 queries composed by Database Administrators (DBAs) for Business Intelligence scenarios combined with four views. 
Compared to Spider~\cite{yu2018spider}'s 10181 questions and BIRD~\cite{li2024can}'s 12751 questions, the SRD and MRD datasets contain too few queries. 
This is mainly because, in actual NL2BI tasks, most questions do not have what is referred to as a ``golden SQL''. 
For example, a user might ask, ``\textit{How many new users were added in the last three days?}'' (The one of real user questions in Baidu Inc).  
The SQL below represents the number of users added each day over the past three days: 
\begin{lstlisting}[ language=SQL,
                    deletekeywords={IDENTITY},
                    deletekeywords={[2]INT},
                    morekeywords={clustered},
                    framesep=8pt,
                    xleftmargin=40pt,
                    framexleftmargin=40pt,
                    frame=tb,
                    framerule=0pt ]
SELECT event_day, COUNT(DISTINCT uid) 
FROM t6628
WHERE is_video_new = 1 
AND event_day BETWEEN DATE_SUB(CURRENT_DATE, INTERVAL 3 DAY) AND DATE_SUB(CURRENT_DATE, INTERVAL 1 DAY)
GROUP BY event_day
ORDER BY event_day ASC;
\end{lstlisting}
, while the following SQL indicates the total number of users added over the same period: 
\begin{lstlisting}[ language=SQL,
                    deletekeywords={IDENTITY},
                    deletekeywords={[2]INT},
                    morekeywords={clustered},
                    framesep=8pt,
                    xleftmargin=40pt,
                    framexleftmargin=40pt,
                    frame=tb,
                    framerule=0pt ]
SELECT COUNT(DISTINCT uid) 
FROM t6628
WHERE is_video_new = 1 
AND event_day BETWEEN DATE_SUB(CURRENT_DATE, INTERVAL 3 DAY) AND DATE_SUB(CURRENT_DATE, INTERVAL 1 DAY);
\end{lstlisting}
. 
Here, \textit{is\_video\_new} represents whether a user is new, with \textit{is\_video\_new} = 1 indicating a new user and \textit{is\_video\_new} = 0 indicating not a new user. 
The \textit{uid} represents the user ID, and \textit{event\_day} represents the date. 
In the BI scenario, it can be assumed that today's data is not yet prepared. 
Thus, similar to the previous work~\cite{floratou2024nl2sql}, we use useful execution accuracy (UEX) rather than execution accuracy (EX) to evaluate, as both SQLs are considered correct in our context, whereas in other datasets, only one result would be deemed correct. 
Expanding the dataset size is hampered by the challenge of accurately assessing effectiveness, which poses a significant challenge.


%% file: 6.related_conclusion.tex
\section{Related Work}

\noindent
\textbf{Prompting-Based NL2SQL. }
LLMs such as GPT-3~\cite{openAIGPT}, LLaMA~\cite{LLama}, Codex~\cite{OpenAICodex}, PaLM~\cite{narang2022pathways}, and Claude 3~\cite{Claude3} are breaking traditional practices in natural language processing (NLP) communities, providing new possibilities for solving NL2SQL tasks. 
Inspired by chain-of-thought~\cite{wei2022chain}, an increasing number of studies focus on designing effective prompting techniques to better utilize LLMs for various subtasks. 
For example, in the NL2SQL scenario, DIN-SQL~\cite{pourreza2024din}, C3~\cite{dong2023c3}, and SQL-PaLM~\cite{sun2023sql} provide prompting techniques to GPT-4, GPT-3.5, and PaLM respectively, greatly enhancing the accuracy of LLMs in generating NL2SQL tasks. 
GPT mentions in the paper that using different data structure forms, such as using the character \# versus not using it, can make a significant difference. 
Therefore, it is necessary to focus on effective and novel data structures for prompting techniques.

\noindent
\textbf{Pre-trained and Supervised Fine-Tuning-Based NL2SQL. }
In the domain of NLP, prior to the emergence of LLMs, the predominant methodology~\cite{sutskever2014sequence, graves2012long, choi2021ryansql, hwang2019comprehensive, wang2019rat, cai2021sadga, cao2021lgesql, gan2021natural, yin2020tabert, herzig2020tapas, yu2020grappa} for addressing the NL2SQL task involved the fine-tuning of an "encoder-decoder" model. 
Leveraging the sequence-to-sequence paradigm, this approach facilitated the modeling of NL2SQL as a transformation from natural language to the corresponding SQL output. 
Furthermore, the advent of pre-training techniques enabled supervised fine-tuning on specific datasets, thereby augmenting the encoder's capabilities and consequently yielding a more effective resolution of the NL2SQL task. 
Typically, this process necessitates a considerable volume of data pertinent to databases and SQL. 
Unlike methodologies predicated upon large language models and prompting strategies, the pre-training of language models in this approach does not accommodate the guidance of thought chains during utilization, rendering it somewhat constrained. 

\noindent
\textbf{LLM Designed Specifically for NL2SQL. }
The recent study \textit{CodeS}~\cite{li2024codes} has developed a LLM tailored specifically for the NL2SQL task, with a parameter size ranging from 1B to 15B, yielding commendable outcomes. 
Nonetheless, it is imperative to take into account both the costs associated with training and data acquisition. 
In comparison to Supervised Fine-Tuning (SFT), this approach necessitates a more extensive dataset. 
Such methodologies exhibit a propensity towards specialization, diverging from the inherent versatility of LLMs (i.e., their adaptability to a multitude of subtasks). 
\textit{CodeS} still relies on high-quality training data, which is often difficult to obtain. 
In the absence of sufficient training data for the NL2BI task, the data augmentation and few-shot inference approach proposed by \textit{CodeS} have not been validated in this scenario. 
Consequently, this particular LLM type falls outside our scope of consideration.

\section{Conclusion}
In this paper, we describe and evaluate ChatBI, a technology that converts Natural Language into complex SQL for Business Intelligence (BI) scenarios. 
First, ChatBI employs smaller and cheaper models compared to LLMs to match the Multi-Round Dialogue scenarios encountered by users in practice. 
Second, ChatBI initially leverages view technology from the database community to transform challenging schema linking and metric ambiguity issues in BI scenarios into a Single View Selection problem, also using smaller and cheaper models to address this problem. 
Finally, ChatBI first introduces a phased process flow, which decompose the NL2BI task. 
ChatBI then proposes \textit{Virtual Columns} mapping to handle complex semantics, computations, and comparative relationships. 
Based on this design, ChatBI effectively understands these complex relationships within the NL2BI task. 
In actual production systems, ChatBI has enabled many non-expert users to perform expert-level BI analysis.